# Advanced Lung Nodule Segmentation and Classification for Early Detection of Lung Cancer using SAM and Transfer Learning

Asha V [a,b], Bhavanishankar K [a,b]

[a]Dept of Computer Science and Engineering, RNS Institute of Technology , Bengaluru, 560098, Karnataka, India
[b] Visvesvaraya Technological University , Belagavi , 590018, Karnataka, India

**Abstract**

Lung cancer is an extremely lethal disease primarily due to its late-stage diagnosis and significant mortality rate, making it the major cause of cancer-related demises globally. Machine Learning (ML) and Convolution Neural network (CNN) based Deep Learning (DL) techniques are primarily used for precise segmentation and classification of cancerous nodules in the CT (Computed Tomography) or MRI images. This study introduces an innovative approach to lung nodule segmentation by utilizing the Segment Anything Model (SAM) combined with transfer learning techniques. Precise segmentation of lung nodules is crucial for the early detection of lung cancer. The proposed method leverages Bounding Box prompts and a vision transformer model to enhance segmentation performance, achieving high accuracy, Dice Similarity Coefficient (DSC) and Intersection over Union (IoU) metrics. The integration of SAM and Transfer Learning significantly improves Computer-Aided Detection (CAD) systems in medical imaging, particularly for lung cancer diagnosis. The findings demonstrate the proposed model effectiveness in precisely segmenting lung nodules from CT scans, underscoring its potential to advance early detection and improve patient care outcomes in lung cancer diagnosis. The results show SAM Model with transfer learning achieving a DSC of 97.08% and an IoU of 95.6%, for segmentation and accuracy of 96.71% for classification indicates that ,its performance is noteworthy compared to existing techniques.

Segment Anything Model (SAM), Transfer Learning, Vision Transformer Model, Bounding Box Prompts, Computer-Aided Detection (CAD), Computed Tomography (CT) scans

## 1. Introduction

Lung cancer is among the most critical type of cancer, marked by a high mortality rate and frequently delayed detection. Globally, it remains the top cause of concern for cancer-related deaths. In 2024, statistics indicate that lung cancer is accountable for approximately 18% of all cancer fatalities, leading to an estimated 1.8 million deaths every year. Figure 1(left) shows lung and bronchus cancer risk with the increase in age, with males generally having a higher risk than females in 2024. Figure 1(right) highlights higher lung and bronchus cancer mortality in males and significant breast cancer deaths in females. These statistics underscore the critical need for early detection and effective treatment strategies to improve patient outcomes [1-2].

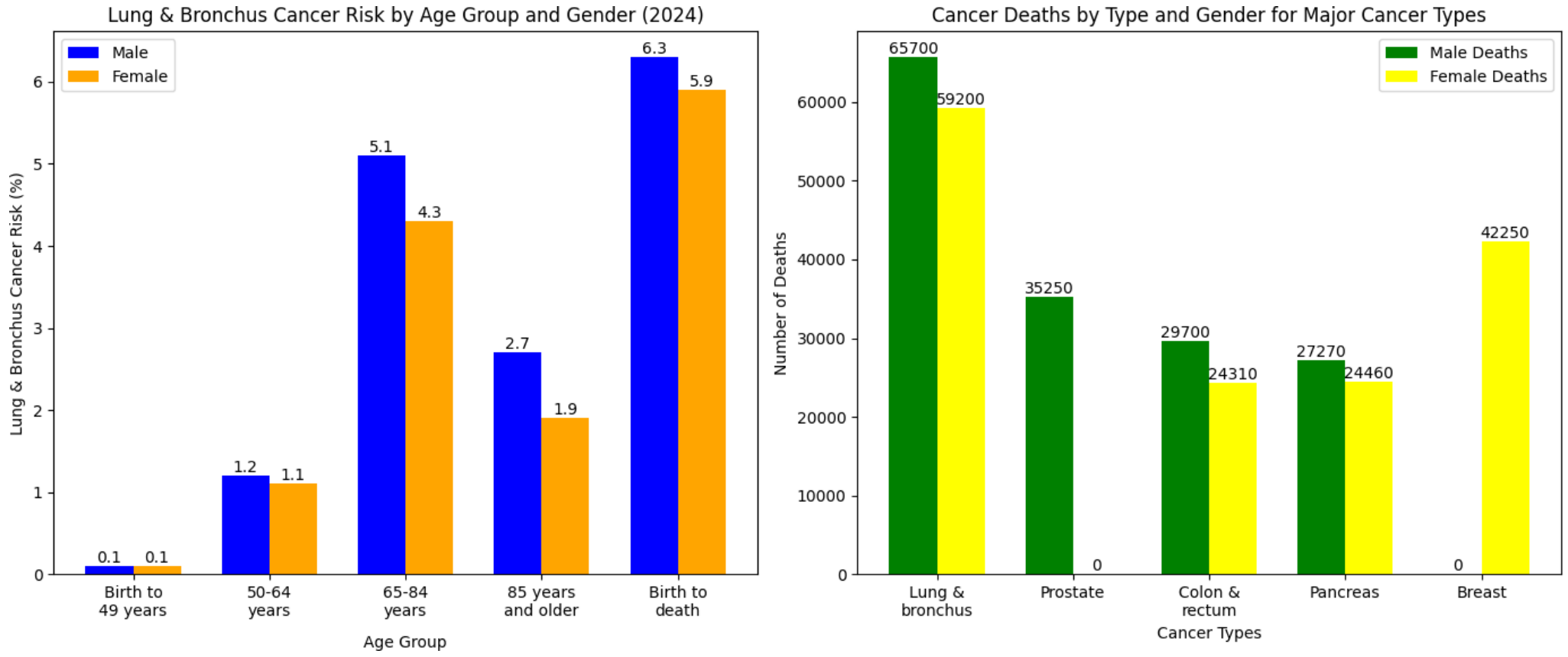

Figure 1: (Left) Lung and Bronchus Cancer Risk by Age Group and Gender (2024) and (right) Major Cancer Types and Gender-Specific Deaths.

Medical imaging techniques are essential for cancer detection, diagnosis, and treatment planning. Widely used modalities include CT, MRI, Positron Emission Tomography (PET), and X-rays [4-7]. CT is especially favored for lung cancer because of its high-resolution images, allowing for detailed visualization of the lung tissues and detection of small nodules [4]. However, the complexity and volume of medical images necessitate CAD systems to assist radiologists to diagnose accurately. The typical CAD application for lung cancer detection is depicted in the Figure 2. CAD systems enhance accuracy, reduce diagnostic times, minimize hu- man error and addressing the limitations of manual analysis [3].

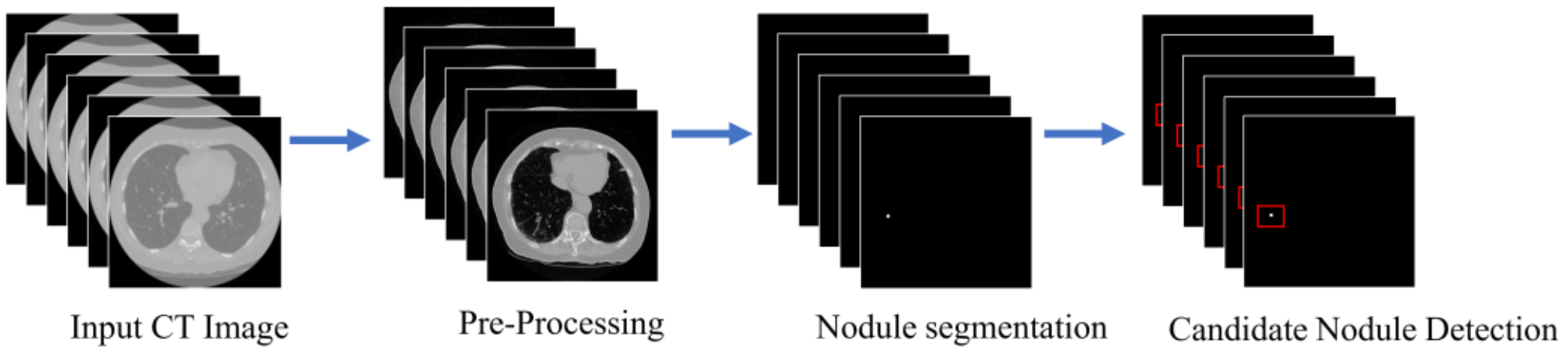

Figure 2: CAD system stages for lung nodule segmentation.

Medical image segmentation, particularly for lung and lung nodule segmentation, plays a virile role in ensuring accurate diagnosis, treatment framework, and real time monitoring of lung cancer [8]. Conventional image processing techniques, such as edge detection, thresholding, and region-based methods, often face limitations in handling the inherent complexity and variability of medical images [8]. ML and DL approaches have gained prominence in medical diagnostics. CNNs, a fundamental DL architecture, are specifically designed to analyze visual data like images and videos, making them exceptionally suited for complex image analysis tasks in healthcare. While 2D CNNs are widely used for image analysis, they have limitations with 3D data, as they cannot fully capture volumetric spatial relationships. To address this, 3D CNNs have been developed, offering enhanced capacity to extract spatial features across depth, width, and height, thus improving performance in applications requiring comprehensive three-dimensional context. Computer Aided Detection (CAD) tools enable automatic lung nodule identification in chest CT scans, essential for early lung cancer diagnosis, enhancing radiologist's efficiency and diagnostic accuracy through stages like preprocessing, segmentation, nodule identification, feature extraction, and classification. Despite advancements, challenges remain, such as acquiring high-quality data, achieving precise segmentation, and obtaining accurate annotations, data imbalance is also a significant issue. DL methods demonstrate potential for improving the accuracy and speed of lung cancer detec-tion, though hurdles like the need for high-quality data and complex image annotation persist. Effective DL models must overcome these challenges, particularly in segmentation and classification phases, where high-resolution thoracic images capture detailed structures that complicate processing and analysis. Reliable DL models capable of learning from clinical records and medical images are essential for early-stage detection, focusing on specific regions of interest. The advent of automatic segmentation techniques like the SAM, manifests a significant advancement. SAM leverages DL to pro- vide precise and efficient segmentation, overcoming limitations of traditional methods and facilitating improved clinical decision-making. Its ability to segment various anatomical structures with minimal user intervention underscores its potential in transforming medical image analysis [5]. Similarly, the classification of lung cancer malignancy plays important role in its diagnosis. Currently, machine learning based methods are adopted widely for medical image analysis for pattern recognition and classification in computer –aided diagnosis (CAD) systems. The machine learning and deep learning-based methods have gained attention in this domain. Several methods have been introduced for lung cancer classification such as Ren et al. [20] developed deep learning based ensemble framework by using LeNet, GoogleNet, AlexNet, ResNet, VGG16 and DenseNet by using histopathological images. In [21] authors introduced CNN and attention based model for classification of lung cancer. Humayun et al. [22] adopted deep transfer learning with CNN approach for lung cancer classification. The classification and segmen- tation of lung nodules has been studied widely in this domain of medical image processing systems. The deep learning methods have proven their significance in these applications however the segmentation and classification accuracy remains challenging issue. To overcome this issue, the focus is on incorporating a novel segmentation approach for lung nodule segmentation with transfer learning and mobilenet for malignancy classification. Recently, Kirillov et al. [23] introduce Segment Anything Model (SAM) for zero shot image segmentation however this model has obtained noteworthy performance for non-medical images. Several researches has adopted this model for segmenting the complex biomedical images, a detailed review on medical image segmentation by using SAM [23] is presented in section II. The complex nature of biomedical images remained a challenging task which has negative impact on the overall segmentation accuracy. The proposed work performs two tasks: lung nodule segmentation by using SAM and lung cancer malignancy classification by using MobileNetV2.

### *1.1. Contributions of the work*

**Innovative Integration of SAM and Transfer Learning:** The study introduces a novel method for lung nodule segmentation by combining the Segment Anything Model (SAM) with transfer learning techniques, enhancing the performance and accuracy of lung nodule detection from CT scans.
**Superior Performance with Bounding Box Prompts and Vision Transformer:** By incorporating Bounding Box Prompts and Vision Transformer model, the proposed method shows enhanced accuracy as compared to the existing benchmark techniques, demonstrating superior segmentation performance, including high accuracy, DSC, and IoU metrics.
**Mobile Net for the malignancy Classification:** Integration of SAM and MobileNetV2 for nodule classification.
This paper is structured in 6 sections, the remaining 5 sections are as follows. Details on prevailing lung nodule CAD systems in the literature are provided in Section 2. Section 3 highlights the existing SAM model and the finetuned SAM model for lung nodule detection system with classification in detail. Section 4 describes details of the dataset used for the proposed work, experimental setup and training procedure. Section 5 presents the outcomes derived from the suggested methodology and compares them with the leading techniques in the field. Section 6 concludes by outlining the future scope of this research.

## 2. Literature Survey

This section surveys prevailing DL methods for lung nodule segmentation. DL gives a significant contribution in the treatment of lung cancer by enabling precise segmentation of lung nodules, thereby improving early detection and assisting in accurate treatment planning. DL models excel in medical im- age segmentation, offering precise results by learning intricate features from medical imaging data. However, their task-specific designs often necessitate generalization for broader clinical applications. Recent advancements in segmentation models have demonstrated improved versatility and performance across diverse tasks [9]. For instance, the development of DEHA-Net, a dual encoder-based architecture, has leveraged CT scans and coronal views [10], enhancing segmentation accuracy with multi-scale and residual structures. The traditional UNet model has also been refined with similar architectural improvements [11]. Moreover, innovations such as MetaAI's training-robust models for zero-shot image data segmentation [12], and SAM's universal models for prompt-based segmentation [13], have demonstrated SAM's superior accuracy across various medical datasets [14]. Adapted MedSAM [15] has further reported improved accuracy in prompt-based segmentation. Table 1 provides a summary of various DL methods for lung nodule segmentation. Lung nodule classification also considered as challenging yet important task for diagnosis of lung cancer. As discussed before, the computer vision and deep learning based methods have gained huge attention in this domain therefore in this subsection ,description is on the existing deep learning lung cancer classification schemes. Saikia et al. [31] adopted transfer learning based approach for classification where pre-trained deep learning model VGG was used along with support vector machine (SVM) and Random Forest (RF) classification to form the hybrid classifier model. Huang et al. [32] introduced self-supervised transfer learning approach to classify the lung nodule malignancy. According to this approach, the first phase includes eliminat- ing redundant noise from slices therefore adaptive slice selection model is presented as pre-processing phase. Next phase implements self-supervised learning approach for pattern learning. Finally, a transfer learning based approach which relies on domain adaption methods is designed to obtain the robust features for classification. Al-Shabi et al. [33] presented Progressive Growing Channel Attentive Non-Local (ProCAN) network architecture for nodule classification. A channel-wise attention mechanism is added to the non-local network module and curriculum learning method is applied to train the model finally the training process is updated progressively to im- prove the learning capability. Mahmood et al. [34] presented CNN based approach by using AlexNet architecture which is improved by modifying the layer ordering, hyperparameters and various functions of the network. More- over, several pre-processing steps are also implemented such as segmentation, normalization and zero centring to improve the learning process. Dodia et al. [20] introduced novel deep-learning architecture, named receptive field regularized V-net (RFR V-Net), is proposed for detecting lung cancer nodules with reduced false positives (FP).

| Ref. | Method | Advantage | Limitations | Dataset Used | Performance Metrics |
|---|---|---|---|---|---|
| [9] | DEHA-Net | Automatic ROI generation, multi-view analysis | Extensive input requirement from radiologists or CAD systems | LIDC/IDRI | DSC: 87.91%, Sensitivity: 90.84%, PPV: 89.56% |
| [10] | SMR-UNet | Enhanced details, fast convergence | Struggles with diverse shapes and small nodules | LIDC, Fourth Affiliated Hospital of Guangxi Medical University | Dice: 0.9187, IoU: 0.8688; Clinical Dice: 0.7785, IoU: 0.6541 |
| [11] | SKV-Net | High precision, lightweight | Balancing accuracy with lightweight design | Lung Nodule Analysis 2016 (LUNA16) | Accuracy improved by 1.3% over V-Net |
| [12] | MedSAM | Generalizes across modalities and cancer types | Limited to modalities and cancers in training set | Internal and external medical datasets | Better accuracy |
| [13] | SAM | Zero-shot segmentation capability | Inferior to specialized models on medical images | Various public medical image datasets | Performance varied; generally lower than U-Net and similar models |
| [14] | Medical SAM Adapter (MSA) | Superior performance on varied tasks | Initial adaptation required | Various medical imaging modalities | Superior to SOTA methods on 19 medical segmentation tasks |
| [15] | SAM for Medical Images | Versatile with manual hints, large-scale analysis | Inconsistent performance across complex cases | COSMOS 1050K dataset | Variable; better with manual hints; DICE improvement with fine-tuning |
| [18] | ProCAN | High AUC, robust to size variation | Complex training due to progressive growing | Public datasets | AUC: 98.05%, Accuracy: 95.28% |
| [19] | SSTL-DA 3D CNN | Captures 3D discriminative features effectively | Requires large, labeled datasets for best performance | LIDC-IDRI | Accuracy: 91.07%, AUC: 95.84% |
| [20] | RFR V-Net + NCNet | Reduced false positives, high sensitivity | Specific to lung cancer nodules | LUNA16 | DSC: 95.01%, IoU: 0.83, Sensitivity: 98.38%, FPs/Scan: 2.3 |

Table 1: Comparison of methods with their advantages, limitations, datasets, and performance metrics.

## 3. Proposed Model

### 3.1. Overview

This section provides the detailed description of proposed SAM assisted segmentation combined transfer learning model for segmentation and classification of lung nodules. The proposed architecture uses several sub-modules therefore. we briefly describe the base concepts of sub-modules in the following sub-section. The first stage is preprocessing, in this stage image mask generation using annotated data is performed. In next stage SAM is applied where the input CT image is processed via image encoder and prompt encoder model is used where Bounding Box prompts are given as input. Based on these inputs, the mask decoder block generates the segmented map. The output of segmented map is given to the MobileNetV2 for the malignancy classification.

### 3.2. The Basic SAM

The SAM model, a revolutionary fundamental paradigm for image segmentation, has recently been published [23]. SAM is based on the vision transformer (ViT) [22] and was trained using a massive dataset that included 1 billion masks and 11 million pictures. Its outstanding zero-shot segmentation accuracy on prior untested datasets and workloads is its most noteworthy characteristic. Several prompts, including boxes and points, which define the target objects' region-level locations and pixel-level meanings, enable this feature. The model has shown to be incredibly flexible, performing a variety of segmentation tasks with ease. Figure 3 depicts the architecture of the SAM model for segmentation.

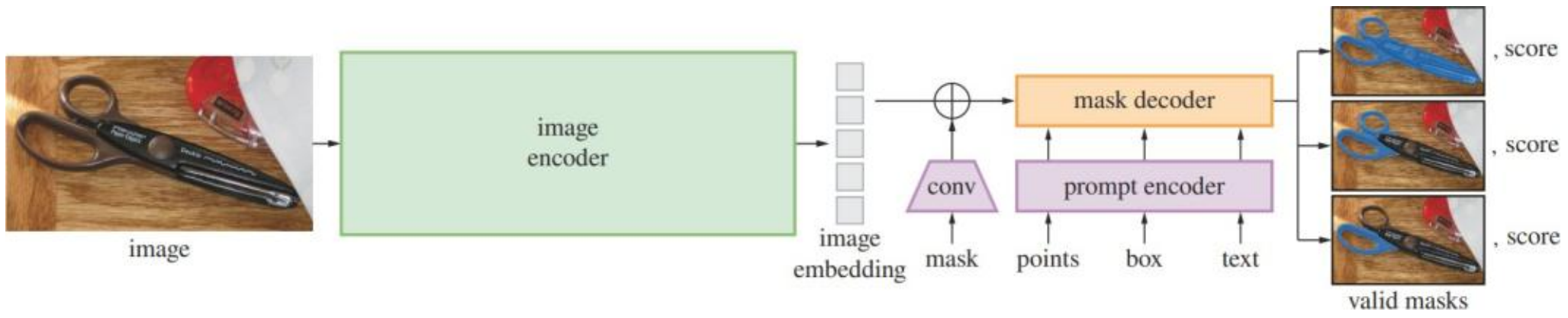

Figure 3: SAM architecture for image segmentation [23].

SAM includes three essential components: an Image Encoder, a versatile Prompt Encoder, and a high-speed Mask Decoder.

]**Image Encoder:**This module uses a ViT, pre-trained using Mask Automatic Encoder (MAE), to handle high resolution inputs. It processes input images sequentially, generating image embeddings. These embeddings, along with their masks, are processed through a convolution layer before moving to the prompt encoder stage.
**Prompt Encoder:** Prompts serve as initial inputs for SAM's segmentation tasks. SAM uses sparse prompts (points, boxes, text) and dense prompts (masks). Sparse prompts are encoded positionally, while mask prompts utilize convolutions and are combined element-wise with image embeddings.
**Mask Decoder:** This module maps image embeddings, prompt embeddings, and output tokens to a mask. Inspired by existing models [22-23], it employs a customized transformer decoder incorporating self-attention and cross-attention blocks is used. Following processing through these blocks, the image embedding undergoes up-sampling, and the result is mapped to a dynamic linear classifier using the Multilayer Perceptron (MLP), which calculates the mask foreground probability for every point in the picture. Transfer learning reuses or adapts a model developed for one task to a related task. It uses information from one issue to tackle another, rather than starting from scratch when training a model. The given below figure 4 represents how SAM segments the 2 different lung CT source images and its corresponding-colored masks. It is very difficult to know which is the actual nodule from the generated masks. Hence, it is required to train the model for the annotated masks to detect the candidate nodule. The below section describes how SAM is tuned for the LUNA16 dataset for the lung nodule detection.

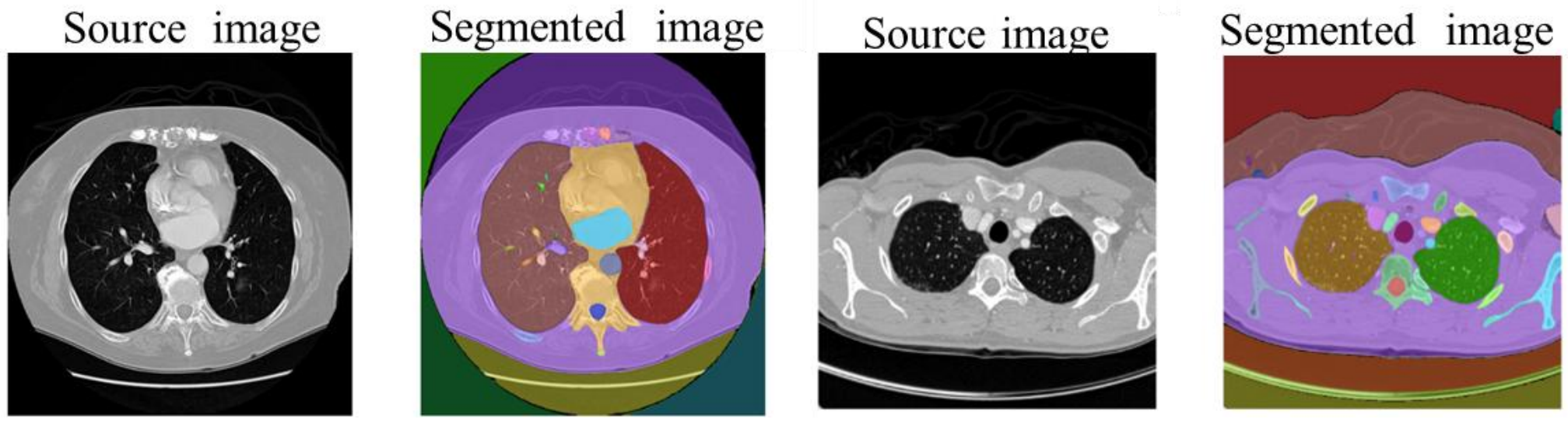

Figure 4: Comparison of Source and Segmented Images using SAM.

### 3.3. *Finetuned SAM for customized LUNA16 dataset to segment the Lung Nodules using Bounding Box prompts*

**Image Encoder:** It is a module that generates image embeddings with dimensions.The dimensions of the feature map are $C \times H \times W$, where:

- $C$ represents the number of channels.
- $H$ represents the height.
- $W$ represents the width.

In this work, a pre-trained ViT model, ViT-H/16 (sam vit_h_4b8939.pth), with $14 \times 14$ windowed attention along with 4 equal-spaced global attention modules is used. The image encoder produces a 16 $\times$ 16 downscaled embedding of the lung image data that was sent in. To simplify the execution process, we have rescaled the images to produce input images with a resolution of 256 $\times$ 256 , resulting in corresponding embeddings of 16 $\times$ 16. Furthermore, this model uses 1 $\times$ 1 and 3 $\times$ 3 convolutions to obtain 256 channels, where each convolution is followed by layer normalization.

**Prompt Encoder:** The input prompt is converted into a vector representation using the simple text encoder known as the Prompt Encoder. It converts the Bounding Box prompt's corner points into 256-dimensional vector embeddings [8]. In particular, an embedding pair made up of the top-left and bottom-right points of the corners represents each Bounding Box.

**Mask Decoder:** The encoder module generates the image embeddings which need to processed as the real-time user interactions therefore to facilitate the embeddings user interaction, the decoder module is employed. As discussed in [23], the decoder module consists of two transform layers which are used to fuse the obtained image embeddings and prompt encodings along with the two transposed convolution layers to enhance the resolution of image. The embedding then passes through a sigmoid activation and is subsequently resized to the input dimensions using bi-linear interpolation. The pre-trained model, with its learned weights and architecture, serves as a starting point. Fine-tuning involves adjusting the pre- trained model using LUNA16 dataset specific to the target task, often modifying or replacing final or intermediate layers. Advantages of transfer learning include reduced training time, improved performance, and increased generalization based on the original task's learned features and patterns. The following figure 5 depicts the overall architecture of the proposed work.

**Transfer Learning:** Transfer learning reuses a pre-trained model, like VGG or ResNet, developed for a specific task, to solve a related task. Instead of training from scratch, it leverages the learned features and representations, fine-tuning the model with new data. This approach reduces training time, improves performance, and enhances generalization by applying knowledge from the original task to the new one.

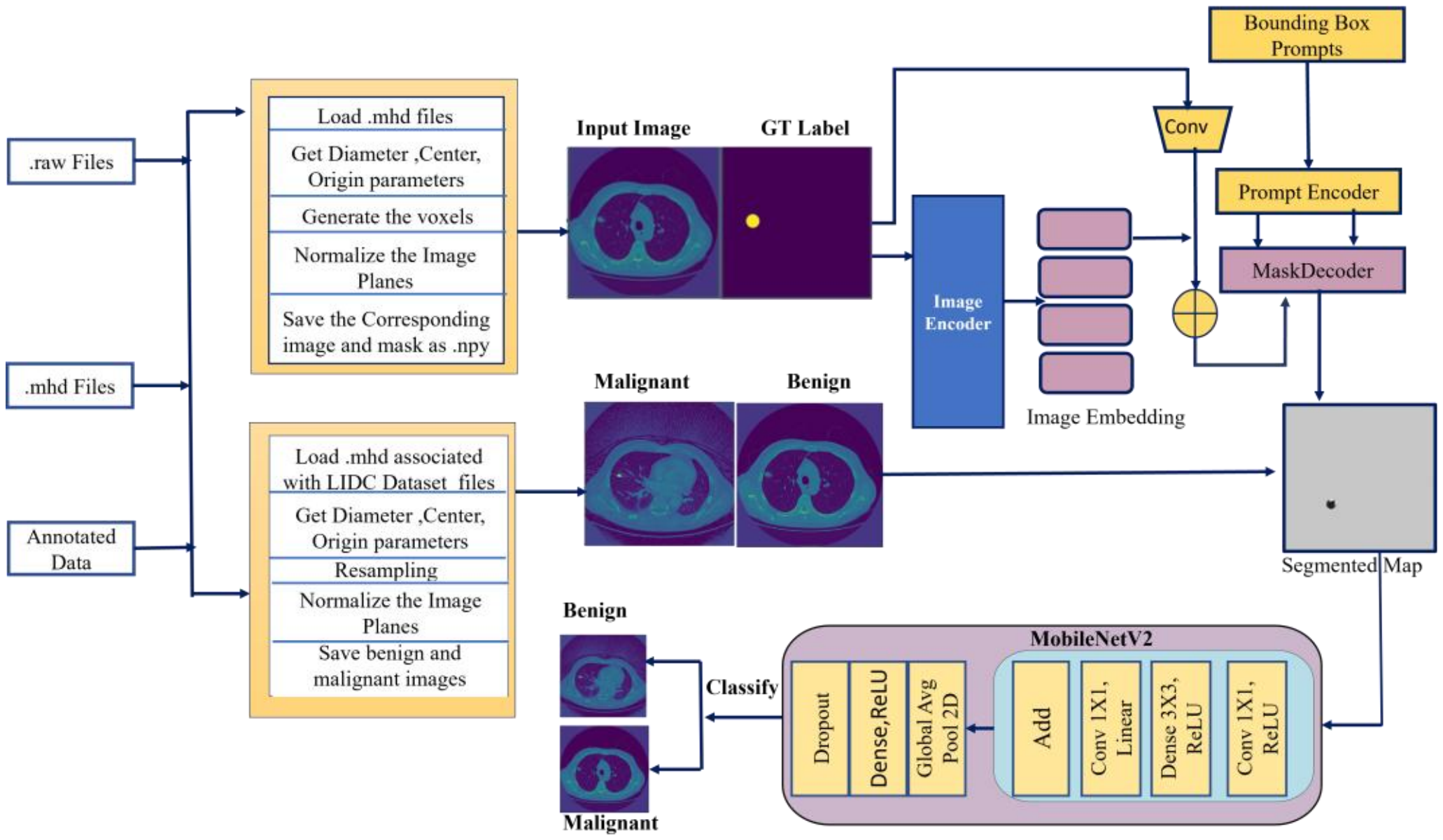

Figure 5: Overall architecture of the proposed work

### *3.4. MobileNetV2 Model for classification of lung cancer*

In this work, our method employs a transfer learning architecture based on MobileNet to classify lung cancer. Convolutional neural network designs in the MobileNet family were created especially for mobile and embedded devices, emphasizing efficiency while preserving a respectable level of accuracy across a range of computer vision tasks. Depthwise separable convolutions, which are made up of a depthwise convolution and a pointwise convolution, are used in the MobileNet architecture. In order to reduce the computational complexity, it is helpful to divide the typical convolution operations into two steps: channel-wise filtering and spatial filtering. Compared to conventional convolutional networks, these depthwise separable convolutions require fewer parameters and calculations, which makes them appropriate for use on devices with constrained processing power. Additionally, by providing parameters like a width multiplier and a resolution reducer, this design enables users to scale the model's width (number of channels) and resolution to strike a balance between efficiency and accuracy, thereby adapting it to various resource restrictions. To increase economy and performance in terms of speed and accuracy for activities like picture classification, this classification uses MobileNetV2, which is an evolution of the MobileNet architecture. This architecture employs the inverted residual technique, in which a lightweight linear bottleneck layer expands the input to a higher-dimensional space. The input is then processed by a depthwise convolution and a pointwise convolution, which projects it back to a lower-dimensional environment. This structure facilitates the effective capturing of more intricate details. By lowering the number of channels in the intermediate bottleneck layer, its linear bottleneck design also lessens the influence on computational efficiency, enabling the network to retain a favorable trade-off between efficiency and performance. Additionally, it incorporates shortcut connections that resemble residual networks (ResNets) in order to improve gradient flow during training and mitigate the issue of disappearing gradients. The given figure 6 depicts the architecture of MobileNetV2.

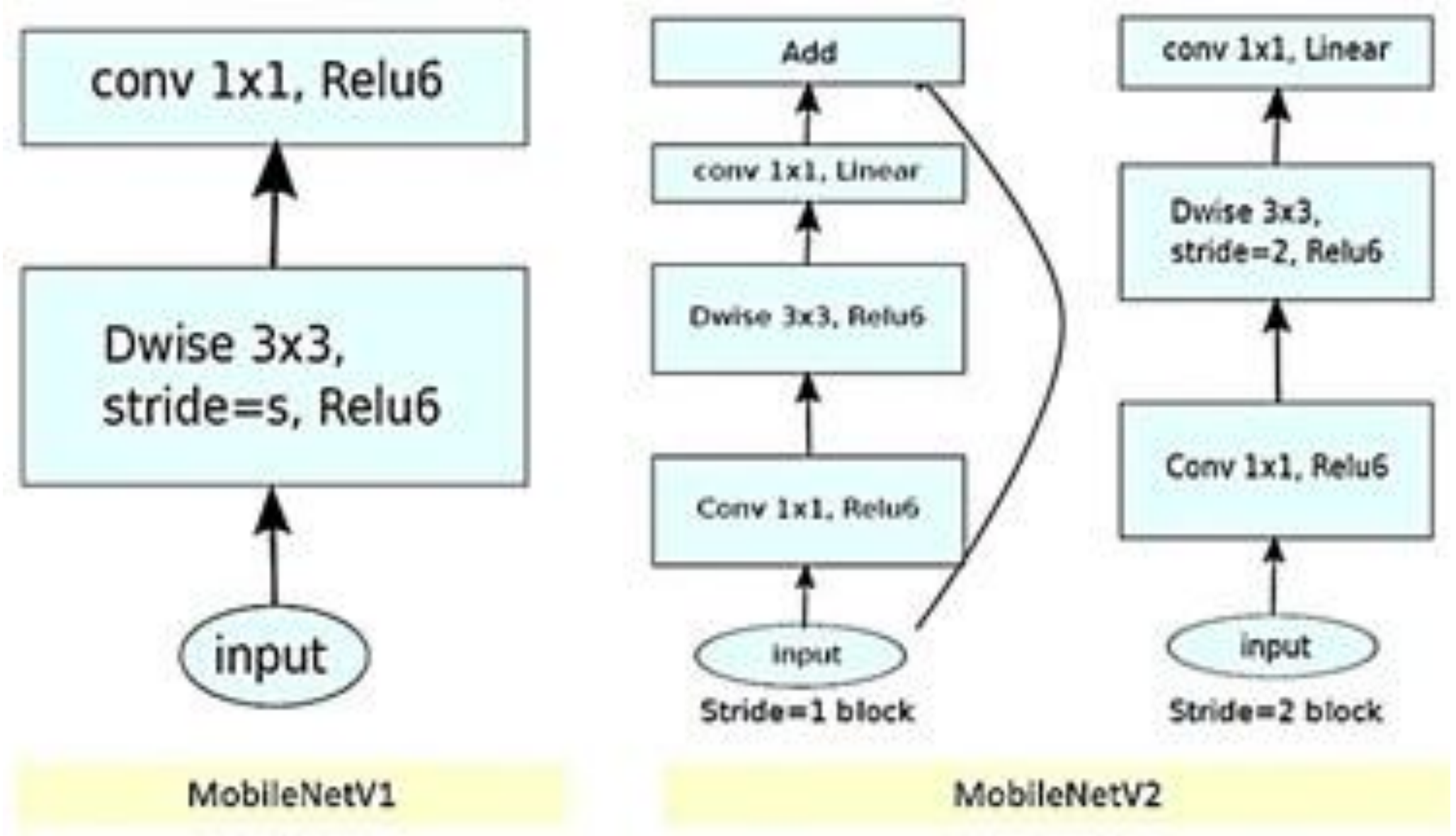

Figure 6: Architectural view of MobileNetV1 and MobileNetV2

| Input | Operator | Output |
|---|---|---|
| $h \times w \times k$ | $1 \times 1$ Conv2D, ReLU6 | $h \times w \times (tk)$ |
| $h \times w \times (tk)$ | $3 \times 3$ Depthwise, ReLU6 | $\frac{h}{s} \times \frac{w}{s} \times (tk)$ |
| $\frac{h}{s} \times \frac{w}{s} \times (tk)$ | Linear $1 \times 1$ Conv2D | $\frac{h}{s} \times \frac{w}{s} \times k'$ |

Table 2: Operations and Transformations

Below given table 3 represents the overall architecture of MobileNetV2. Where t represents expansion factor, c is the number of output channels, n denotes repeating number, s represents strides and this architecture uses 3×3 kernels for spatial convolution.

| Input | Operator | $t$ | $c$ | $n$ | $s$ |
|---|---|---|---|---|---|
| $224^2 \times 3$ | conv2d | - | 32 | 1 | 2 |
| $112^2 \times 32$ | bottleneck | 1 | 16 | 1 | 1 |
| $112^2 \times 16$ | bottleneck | 6 | 24 | 2 | 2 |
| $56^2 \times 24$ | bottleneck | 6 | 32 | 3 | 2 |
| $28^2 \times 32$ | bottleneck | 6 | 64 | 4 | 2 |
| $14^2 \times 64$ | bottleneck | 6 | 96 | 3 | 1 |
| $14^2 \times 96$ | bottleneck | 6 | 160 | 3 | 2 |
| $7^2 \times 160$ | bottleneck | 6 | 320 | 1 | 1 |
| $7^2 \times 320$ | conv2d $1 \times 1$ | - | 1280 | 1 | 1 |
| $7^2 \times 1280$ | avgpool $7 \times 7$ | - | - | 1 | - |
| $1 \times 1 \times 1280$ | conv2d $1 \times 1$ | - | $k$ | - | - |

Table 3: Table of Model Architecture

## 4. Materials and Methods

### *4.1. Dataset formulation and processing for SAM*

The LUNA16 dataset is extensively used in medical imaging and it is curated data from LIDC-IDRI, particularly for CAD systems in lung cancer detection. The original LIDC-IDRI dataset is used for the malignancy values

| Subset ID | Images Count | Min Slices per Image | Max Slices per Image |
|---|---|---|---|
| 0 | 89 | 109 | 733 |
| 1 | 89 | 103 | 636 |
| 2 | 89 | 95 | 682 |
| 3 | 89 | 107 | 764 |
| 4 | 89 | 109 | 564 |
| 5 | 89 | 109 | 633 |
| 6 | 89 | 117 | 590 |
| 7 | 89 | 117 | 522 |
| 8 | 88 | 111 | 580 |
| 9 | 88 | 113 | 567 |

Table 4: Distribution of Image Slices Across LUNA16 Dataset Subsets.

It contains chest CT scans annotated with the presence or absence of lung nodules, potential lung cancer indicators. Released as part of the LUNA16 Challenge by the "Medical Image Computing and Computer Assisted Intervention Society (MICCAI)", the dataset includes 888 CT scans in MetaImage (.mhd) format. Annotations provide 3D coordinates of nodule centroids and their diameters. Table 3 contains distribution of image slices across 9 subsets of LUNA16 dataset. The dataset is stored in. mhd which needs to be converted into suitable format which can be used later for processing. Therefore, we have performed pre-processing phase to produce the corresponding. npy files which are later converted into .jpg format to make it simpler for processing. As shown in Algorithm 1, using nodule mask generation process, lung image and its corresponding ground truth for segmentation is extracted.

### *4.2. Experimental Setup*

The complete model is simulated by using Python programming language with the help of Google Colab and Kaggle web applications. The initial stages include data conversion tasks which are done on local machine which has python 3.8 installed on Windows 11 Pro Operating System. This system is equipped with 64 GB of RAM, and 6 GB of NVIDIA Graphics card. This work uses Adam Optimizer with the Learning Rate (LR) of 0.001, batch size 4 and 100 epochs.

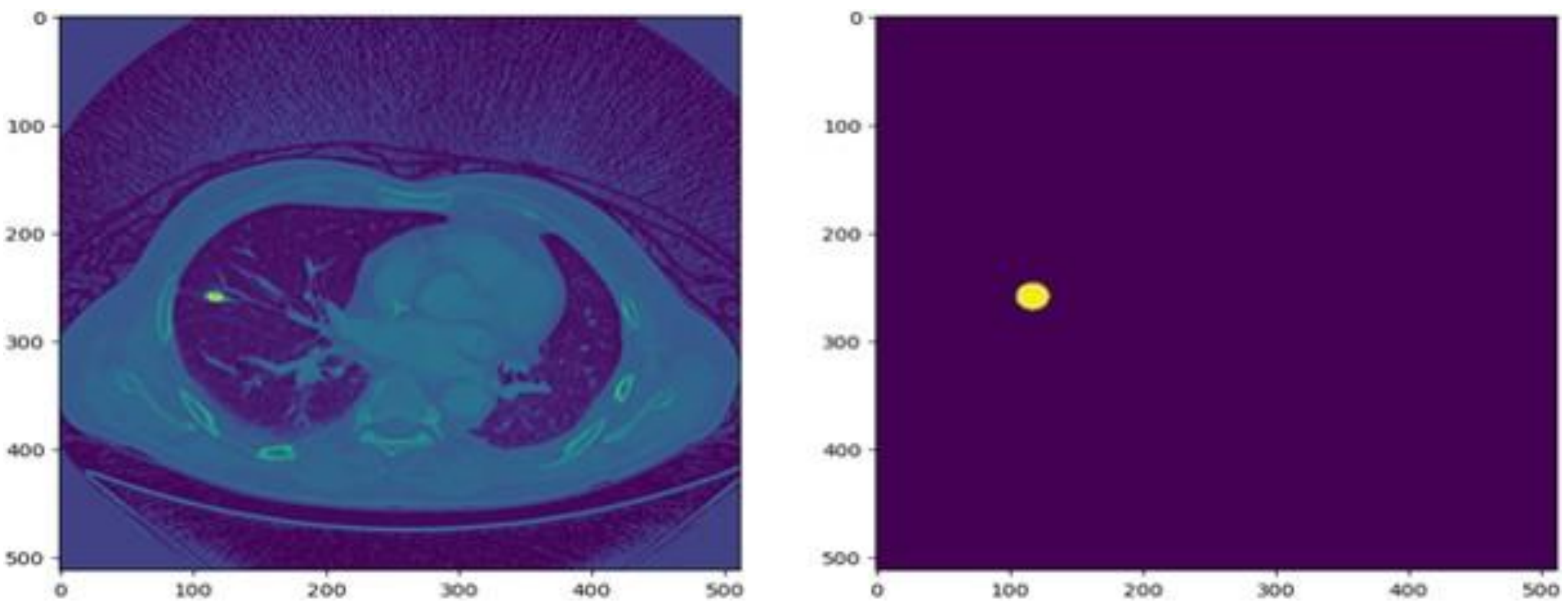

Figure 7: Sample image and its corresponding GT mask of LUNA16 dataset.

### *4.3. Training procedure and loss function*

LUNA16 dataset is utilized here, which has subset 0 to subset 9, where each subset 1-7 has 89 .mhd and their corresponding .raw files and subset 8-9 has 88.mhd images and their corresponding .raw files. For training purpose, we have split the complete data into train and test group where 70% images are used for training and rest 30% images are used for testing purpose. During training process of segmentation module, the loss function is obtained as unweighted sum between Dice loss

$$L_{\text{Dice}}$$

**Algorithm 1** Lung Nodule Mask Generation

**Initialize** the mask generation process
Create an empty mask array:

$$\text{mask}(i, j) = 0 \ \ \forall\, i \in [1, \text{height}],\ j \in [1, \text{width}]$$

Compute Parameters:
Compute the voxel center coordinates:

$$V_{\text{center}} = \frac{x_c - o_x}{s_x}, \frac{y_c - o_y}{s_y}, \frac{z_c - o_z}{s_z}$$

Compute the bounding box for the mask:

$$V_{\text{xmin}} = \max\left(0, \lfloor V_{\text{center}}[0] - V_{\text{diam}} \rfloor - 5\right)$$

$$V_{\text{xmax}} = \min\left(\text{Width} - 1, \lfloor V_{\text{center}}[0] + V_{\text{diam}} \rfloor + 5\right)$$

$$V_{\text{ymin}} = \max\left(0, \lfloor V_{\text{center}}[1] - V_{\text{diam}} \rfloor - 5\right)$$

$$V_{\text{ymax}} = \min\left(\text{Height} - 1, \lfloor V_{\text{center}}[1] + V_{\text{diam}} \rfloor + 5\right)$$

Generate the mask based on these parameters:
**for** each $V_x \in [V_{\text{xmin}}, V_{\text{xmax}}]$ and $V_y \in [V_{\text{ymin}}, V_{\text{ymax}}]$ **do**
    Compute the physical coordinates:

$$p_x = s_x \cdot v_x + o_x, \quad p_y = s_y \cdot v_y + o_y$$

    **if** $\|\text{center} - (p_x, p_y, z)\| \leq \text{diam}$ **then**
        mask $\lfloor \frac{p_y - o_y}{s_y} \rfloor, \lfloor \frac{p_z - o_z}{s_z} \rfloor = 1$
    **end if**
**end for**
Generate PNG images from .npy files:
Save the mask array as a .npy file.
Convert the .npy file to a PNG image.

and cross-entropy loss and GT. Let S be the outcome of the segmentation and G be the corresponding GT where

$$s_i$$

denotes the predicted segmentation map and

$$g_i$$

represents the corresponding ground truth of the voxel i where total N voxels are present in image I. Then, the cross-entropy loss

$$L_{CE}$$

can be expressed as in the Eq. (1)

$$L_{CE} = -\frac{1}{N}\sum_{i=1}^{N} g_i \log(s_i) \tag{1}$$

Similarly, the Dice loss can be represented as:

$$L_{Dice} = 1 - \frac{2\sum_{i=1}^{N} g_i s_i}{\sum_{i=1}^{N} g_i^2 + \sum_{i=1}^{N} s_i^2} \tag{2}$$

With the help of these, the final loss $L$ can be expressed as:

$$L = L_{CE} + L_{Dice} \tag{3}$$

### 4.4. Performance Metrics

The segmentation accuracy of the proposed SAM model is evaluated using several key metrics, including the DSC, IoU, sensitivity (SEN), and Positive Predictive Value (PPV). DSC and IoU serve as the main metrics, measuring the overlap between two segmentation outputs. SEN and PPV, on the other hand, provide additional validation of the model's robustness. In the equations presented below, TP denotes True Positives, FP represents False Positives, TN stands for True Negatives, and FN indicates False Negatives. Table 5 provides the specific formulas used to calculate these metrics. The classification accuracy is measured by using confusion matrix.

| **Performance Metrics** | **Formula** |
|---|---|
| DSC (Dice Similarity Coefficient) | $DSC = \frac{2\times TP}{(2\times TP)+FP+FN}$ |
| IoU (Intersection over Union) | $IoU = \frac{TP}{TP+FN+FP}$ |
| SEN (Sensitivity) | $SEN = \frac{TP}{TP+FN}$ |
| PPV (Positive Predictive Value) | $PPV = \frac{TP}{TP+FP}$ |

Table 5: Evaluation Metrics for Segmentation Task

## 5. Results and Discussion

The outcome of suggested approach is demonstrated here with the comparative analysis against existing methods relevant to medical image segmentation. Figure 8 depicts the SAM for the lung segmentation with a mean loss of 0.0167 and Figure 7 depicts the input image, its corresponding GT and obtained predicted mask. Figure 9 depicts the segmentation outcome using proposed SAM.

### *5.1. Comparative analysis of segmentation*

The obtained performance is evaluated against standard image segmentation methods such as UNet, VNet, FCNUNet, Mask RCNN and EFCM as discussed in [24]. The proposed approach has reported the average performance as 97.08%, 95.6%, 97.85% and 98.1% in terms of DSC, IoU, SEN, and PPV. Given figure 10 depicts the comparative analysis in terms of DSC, IoU, sensitivity(SEN), and positive predictive value (PPV). Figure 11 illustrates the effectiveness of the model training process, showing how both training and validation losses decrease and stabilize over epochs. As the number of epochs increases, the training and validation loss reduce, demonstrating the effective learning of the model and progressive improvement. The training loss generally decreases more quickly than the validation loss. After a certain number of epochs, both losses converge to very low values, suggesting the model has learned the data well and is not overfitting. In next stage, study on nodule malignancy classification study by using label characteristics of LIDC dataset. this dataset contains 9 different semantic features namely Malignancy, Margin, Sphericity, Subtlety, Spiculation, Texture, Calcification, Internal structure and Lobulation. Below given table 6 description of these features and their corresponding ratings which are useful in identifying the class of nodule malignancy.

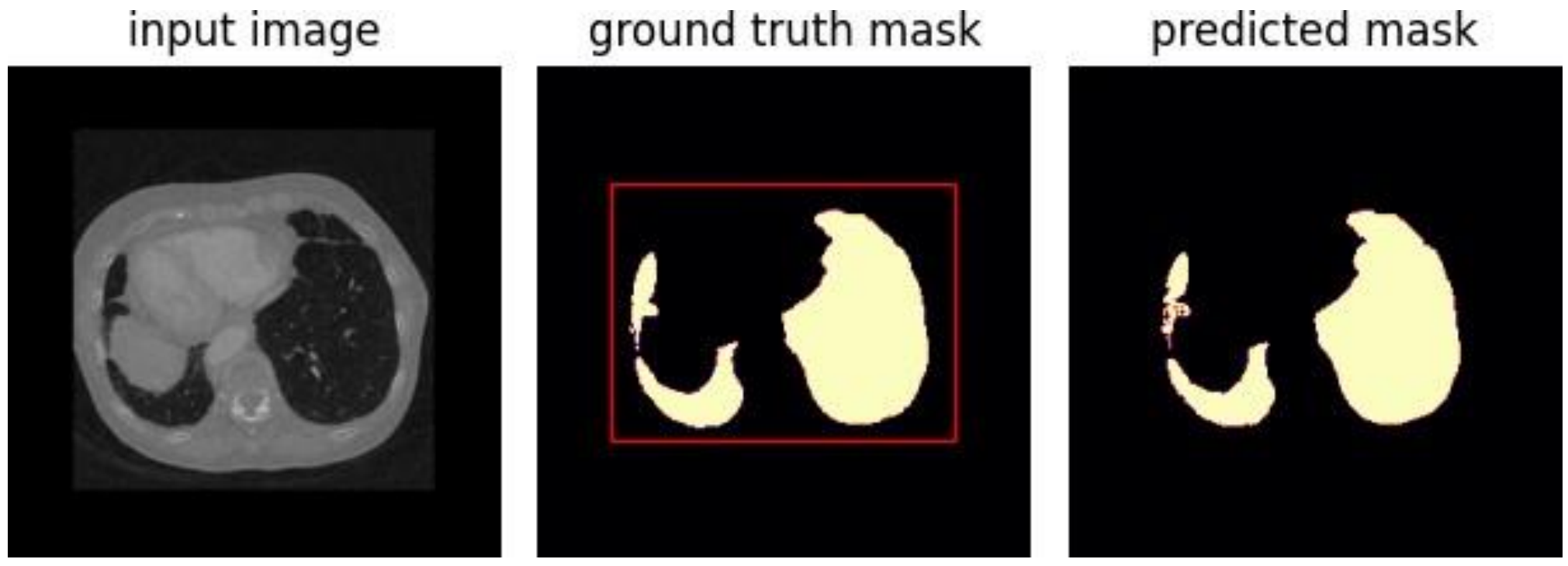

Figure 8: Lung ROI Segmentation using SAM [17].

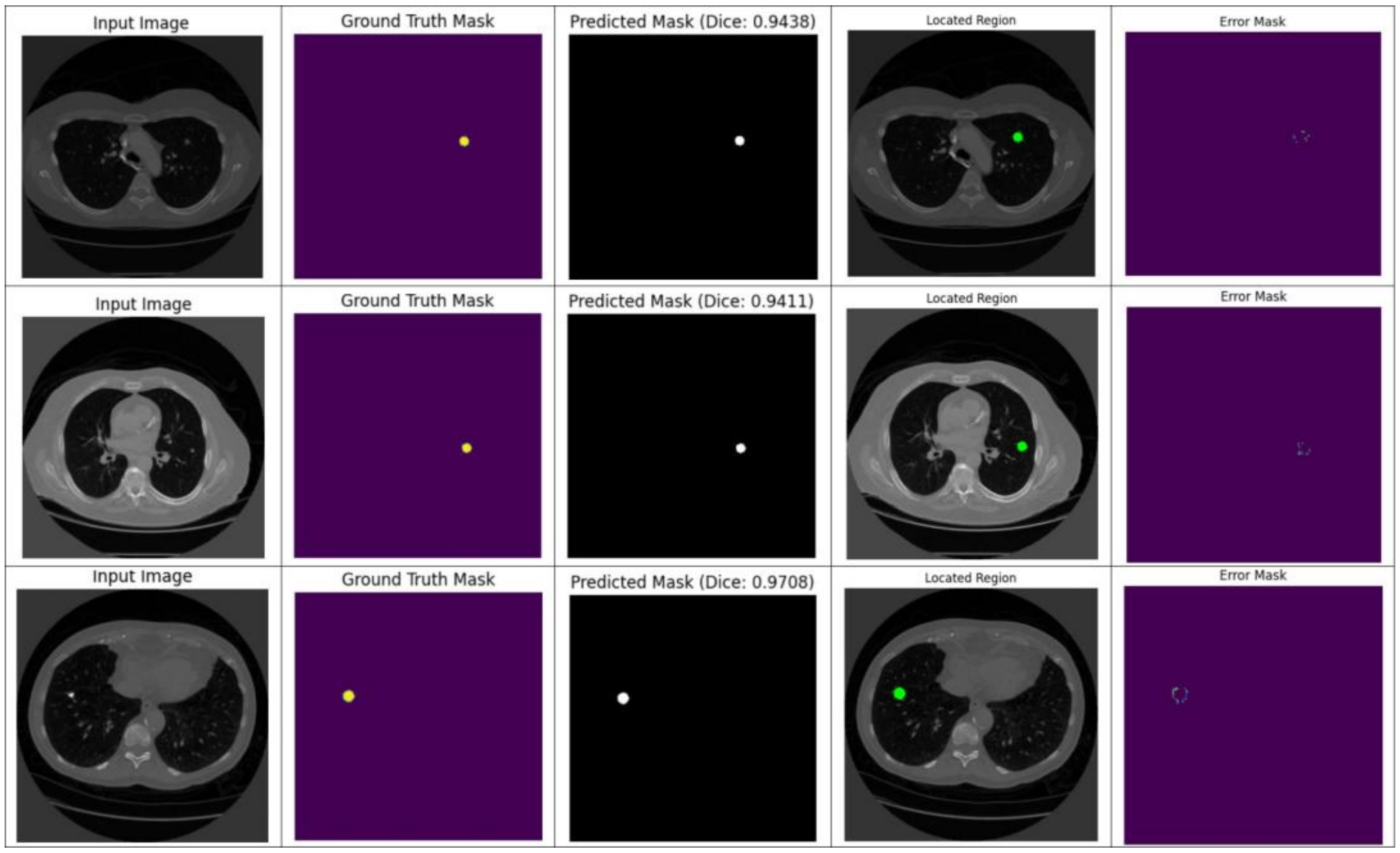

Figure 9: Segmentation outcome (a) Sample input image, (b) Ground Truth, (c) Predicted mask with dice score(d) Located region and (e) Predicted error mask.

| **Nodule Annotations** | **Description** | **Category 0 (Benign)** | **Category 1 (Malignant)** | **Ratings as given in the annotations** |
|---|---|---|---|---|
| Subtlety | Ease of detection relative to surroundings | ≤ 3 Faint visibility | > 3 Prominent visibility | 1. Extremely subtle<br>2. Moderate<br>3. Fairly subtle<br>4. Moderately obvious<br>5. Obvious |
| Internal Structure | Internal composition of the nodule | - | - | 1. Soft tissue<br>2. Fluid<br>3. Air |
| Calcification | Presence and pattern of calcification | ≤ 5 Calcification observed | 6 No calcification present | 1. Popcorn<br>2. Laminated<br>3. Solid<br>4. Non-central<br>5. Central<br>6. Absent |
| Sphericity | Three-dimensional shape in terms of roundness | ≤ 3 Less spherical in shape | > 3 Highly spherical | 1. Linear<br>2. Ovoid<br>3. Ovoid/Round<br>4. Round |
| Margin | Definition of nodule edges | ≤ 3 Indistinct edges | > 3 Clearly defined edges | 1. Poorly defined<br>2. Medium Margin<br>3. Near Sharp<br>4. Sharp |
| Lobulation | Presence and degree of lobulation | - | - | 1. No lobulation<br>2. Nearly no lobulation<br>3. Medium lobulation<br>4. Near marked lobulation |
| Spiculation | Presence and degree of spiculation | - | - | 1. No spiculation<br>2. Nearly no spiculation<br>3. Medium spiculation<br>4. Near marked spiculation<br>5. Marked spiculation |
| Texture | Density of the nodule | ≤ 3 Typically, benign | > 3 Potentially malignant | 1. Non-solid<br>2. Part solid/Mixed<br>3. Solid/Mixed<br>4. Solid |
| Malignancy | Likelihood of malignancy | ≤ 3 Typically, benign | > 3 Potentially malignant | 1. Highly unlikely<br>2. Moderately unlikely<br>3. Indeterminate<br>4. Moderately suspicious<br>5. Highly suspicious |

Table 6: Nodule Annotations and Descriptions

Scores which are ≤ 3 were categorized as 0, whereas scores > 3 were categorized as 1. Category 0 generally represented characteristics of a benign nodule, such as poorly defined margins, lower roundness, limited conspicuity between the nodule and its surroundings, and a non-solid (groundglass-like) consistency. On the other hand, Category 1 tended to indicate features of a malignant nodule, including sharp margins, greater sphericity, high conspicuity between the nodule and its surroundings, and a solid consistency. Below given table 6 shows the final labeled data. Based on these attributes, we train the deep learning model where the entire dataset is divided in 70%-30% train –test ratio. The model is trained for 100 epochs with a batch size of 5. The proposed classification model has reported the overall classification accuracy as 96.71% with loss 0.059. the performance of proposed classification model is compared with the state-of-art deep learning based classification method. Table 8 demonstrates the comparative performance analysis in terms of accuracy, precision, sensitivity, specificity and F1-score.

## 6. Conclusion

This study presented an innovative approach to lung nodule segmentation and classification using the SAM combined with transfer learning. By leveraging the robust capabilities of SAM for image segmentation, the proposed model demonstrated superior performance in segmenting lung nodules from CT scans. The combination of bounding box prompts and a ViT model facilitated accurate and efficient segmentation, which is critical for early lung cancer identification. The output results as shown in the table 7 demonstrates that the suggested approach is better compared to the other standard techniques, achieving high accuracy, DSC, and IoU metrics. This underscores SAM's potential to significantly enhance CAD systems in medical imaging, particularly for lung cancer detection and diagnosis. Although SAM performs well on the LUNA16 dataset, it does not generate accurate masks for all lung CT images. Therefore, further work is needed to improve mask generation accuracy across all lung CT images.

| Techniques | Dice Score | Techniques | IoU |
|---|---|---|---|
| Unet [25] | 94.97% | Pulmonary Nodule Segmentation [26] | 58.00% |
| Dual Branch Residual Network [27] | 82.74% | iWNet [28] | 55.00% |
| Deep FCN [29] | 93.00% | Regression Neural Network [30] | 74.00% |
| Central Focused CNN [31] | 82.15% | Central Focused CNN [31] | 71.00% |
| RFRVNet [20] | 95.01% | RFRVNet [20] | 83.00% |
| EFCM [24] | 97.10% | EFCM [24] | 91.96% |
| **Proposed SAM** | **97.08%** | **Proposed SAM** | **95.60%** |

Table 7: Comparison of Techniques Based on Dice Score and IoU

| Classification Method | Accuracy% | Precision% | Sensitivity% | Specificity% | F1-score% |
|---|---|---|---|---|---|
| Inception V3 | 91.40 | 89.95 | 92.45 | 91.97 | 92.31 |
| ResNet | 87.23 | 86.31 | 89.46 | 87.62 | 88.43 |
| VGG16 | 89.49 | 90.31 | 90.21 | 87.13 | 87.92 |
| DenseNet | 84.21 | 85.42 | 86.68 | 84.96 | 84.57 |
| AlexNet | 90.24 | 91.31 | 89.93 | 90.28 | 91.20 |
| DenseAlexNet | 95.65 | 94.10 | 97.11 | 94.28 | 95.58 |
| **Proposed Model** | **96.71** | **95.25** | **98.30** | **95.45** | **96.50** |

Table 8: Comparison of Classification Methods Based on Evaluation Metrics

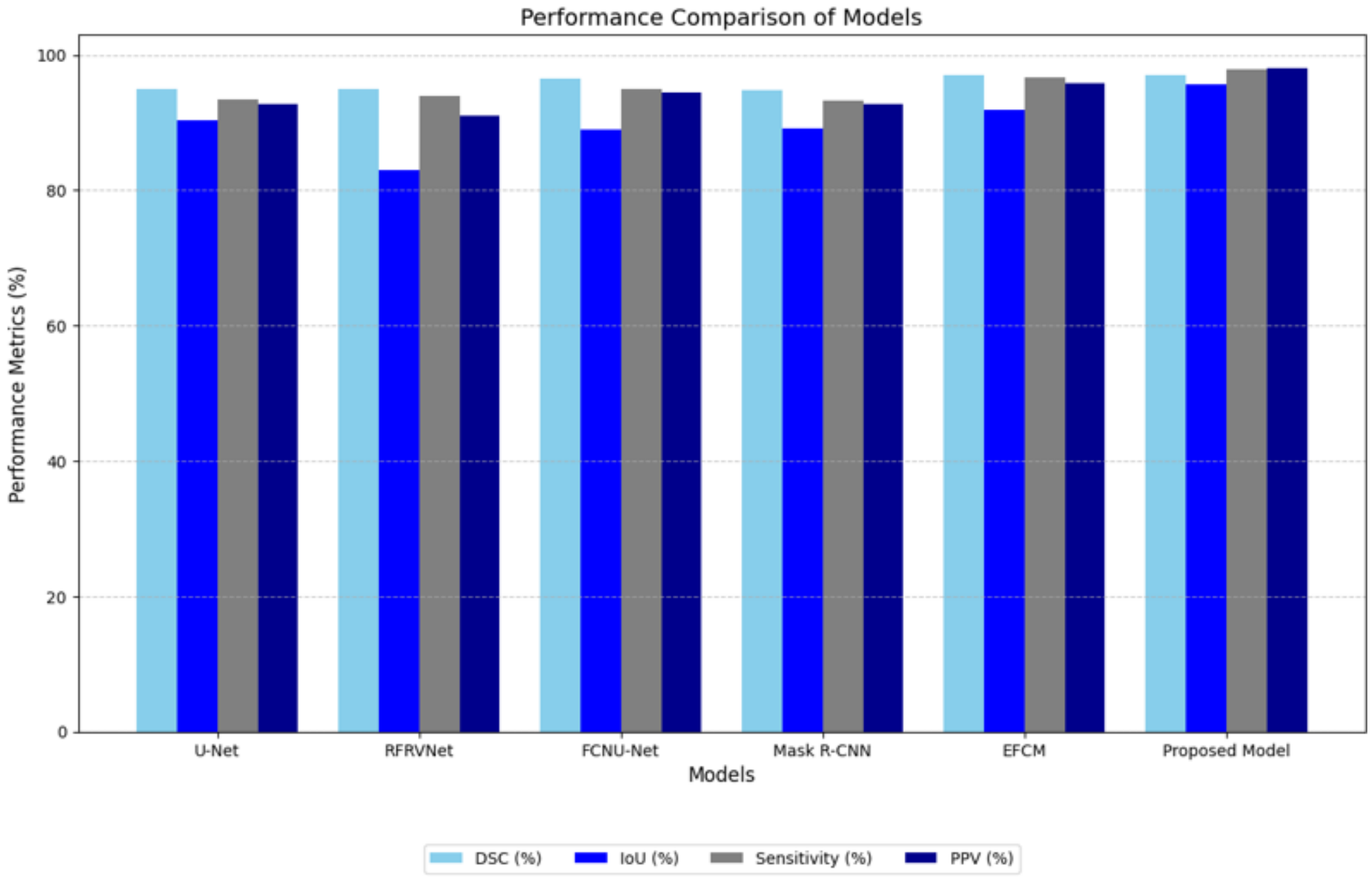

Figure 10: Comparative analysis of Nodule Segmentation

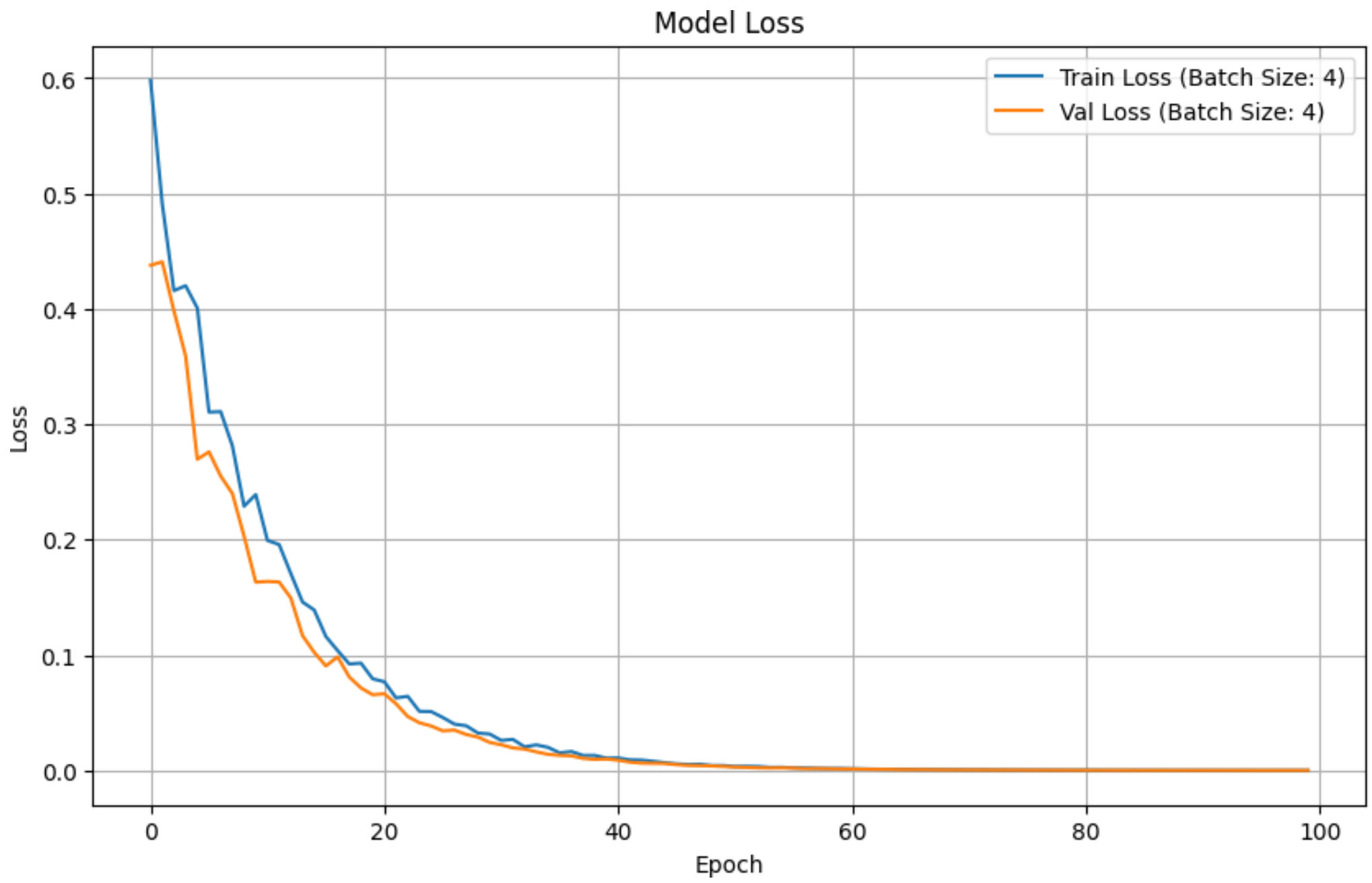

Figure 11: Train and validation loss performance (batch size=4).